\documentclass[aps,prb,twocolumn,showpacs,floatfix,groupedaddress,superscriptaddress]{revtex4-2}
\pdfoutput=1
\usepackage[linkcolor=blue,colorlinks=true,breaklinks=true,citecolor=blue,urlcolor=blue]{hyperref}
\usepackage{graphicx,graphics} 
\usepackage{physics,amsmath,amssymb,amsbsy}
\usepackage{braket}
\usepackage{bm}
\usepackage{dcolumn}
\usepackage{dsfont}
\usepackage[mathscr]{euscript}
\usepackage{xcolor}
\usepackage[titletoc,title]{appendix}
\usepackage[dotinlabels]{titletoc}
\usepackage[caption=false,labelformat=simple]{subfig}

\begin{document}
\title{Gate and Carriers tunable Valley Imbalance in Topological Proximitized Rhombohedral Trilayer Graphene
}
\author{Sovan Ghosh}
\email{sovanghosh2014@gmail.com}
\affiliation{Department of Physical Sciences, Indian Institute of Science Education and Research Kolkata\\ Mohanpur-741246, West Bengal, India}
\author{Bheema Lingam Chittari}
\email{bheemalingam@iiserkol.ac.in}
\affiliation{Department of Physical Sciences, Indian Institute of Science Education and Research Kolkata\\ Mohanpur-741246, West Bengal, India}

\begin{abstract}
We investigated the electronic structure, Fermi surface topology and the emergence of valley imbalance in rhombohedral trilayer graphene (RTG) induced by the topological proximity and the electric fields. We show that, a strong proximity strength isolates the unperturbed low energy bands at the charge neutrality and the isolated topological bands show metallic nature under the influence of applied electric fields. Our calculations indicate that valley-resolved metallic states with a finite Chern number $|C| =$3 can appear near charge neutrality for appropriate electric fields and second-nearest-neighbor strengths. The Fermi surface topology of these metallic bands greatly influenced by the applied electric fields and carrier doping. The valley imbalance lead to the dominant carriers of either $e^-$ or $h^+$ Fermi surface pockets and the choice of carriers is subjected to the direction of electric fields. The gate-tunable and carrier-induced valley imbalance in topologically proximated rhombohedral trilayer graphene may have potential applications toward the realization of superconductivity.             
\end{abstract}

\maketitle
\section{\label{sec:1}Introduction}
Recent work has demonstrated the quantized Hall resistance at zero magnetic field in rhombohedral graphene due to spin-orbit proximity which has opened new avenues for topological band engineering~\cite{sci_hall,ABCA_hall}. Rhombohedral multilayer graphene, particularly RTG, has emerged as a versatile platform for exploring rich electronic phenomena, including strong correlations, symmetry-breaking states, spin and orbital magnetism, and superconductivity~\cite{Arp2024, Zhou2022, Zhou2021, Zhou2021nat}.
For example, aligning rhombohedral stacked graphene with hBN can create flat bands, a precursor to correlated phases \cite{Han2024}. Recent studies have shown that coupling RTG with spin-orbit coupling can lead to exotic phases like valley-XY nematic, superconductivity, and spin-polarized half metal~\cite{PhysRevB.107.L121405,Zhou2021}. External perturbations like proximity-induced spin-orbit coupling through TMDs, can further modify the low-energy band structure of RTG, leading to spin polarization and inter valley coherence states~\cite{PhysRevB.109.035113, ABC_CGT, WS2_AB_CGT, AB_WS2_CGT, AB-WSe2,Wang2015}. The interplay of carrier doping and interlayer potential difference can induce low-energy band condensation, magnetic and Lifshitz transitions, and modify Fermi surfaces, potentially supporting spin-orbit-enhanced superconductivity~\cite{PhysRevB.109.035113}. Recently, it is identified that presence of a single Fermi surface lead to spontaneous spin and valley-isospin symmetry breaking~\cite{Arp2024}. Moreover, the topological proximity effects have been shown to induce quantum Hall phases in graphene~\cite{PhysRevLett.112.096802, PhysRevB.100.081107}. In our recent study~\cite{Sovan_bilayer}, we present such Orbital Hall phases in graphene aligned on a Haldane layer are purely due to the topological proximity.  In this paper, we investigates the topological transitions induced by the electric field and next-nearest-neighbor interactions. Further, we have demonstrated the valley imbalance and the Fermi surface reconstruction in RTG/Haldane heterostructures under influence of interlayer bias and proximity strength.
\begin{figure}
    \centering
    \includegraphics[width=0.8\columnwidth]{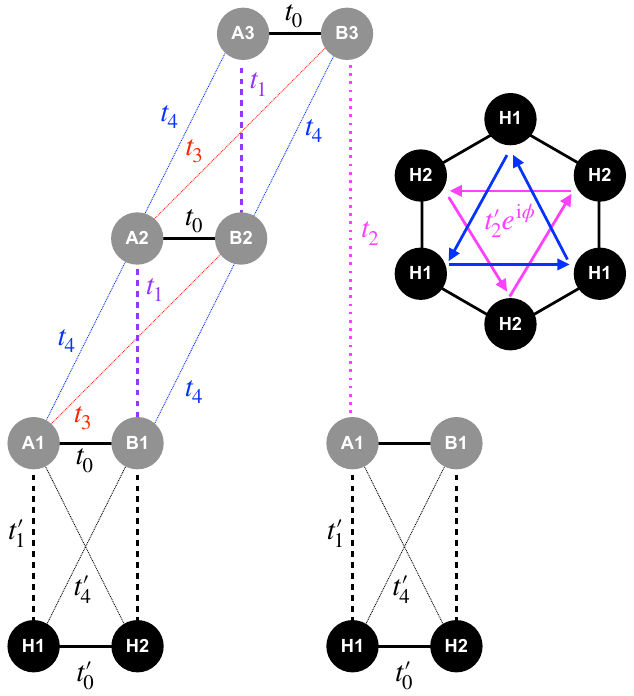}
    \caption{A schematic diagram of rhombohedral trilayer graphene (RTG) aligned on a Haldane layer ($\rm RTG/\tilde{\rm G}$). The gray-filled circles labeled (A1, B1), (A2, B2), and (A3, B3) represent the two sublattices of each graphene layer in RTG. The integer index of each sublattice denotes the layer index. Intra-layer hopping within RTG is represented by $t_0$, while inter-layer coupling is represented by $t_i$ ($i = 1, 2, 3, 4$). The black-filled circles H1 and H2 represent the two distinct sublattices of the Haldane layer. Intra-layer hopping within the Haldane layer is represented by $t^{\prime}_0$, and inter-layer hopping between the Haldane layer and the proximity layer of RTG is represented by $t^{\prime}_1$ and $t^{\prime}_4$. The next-nearest-neighbor (NNN) complex hopping process in the Haldane layer is represented by $t_2^{\prime}e^{i\phi}$.}
    \label{Fig1}
\end{figure}
%
\section{\label{sec:2}Hamiltonian}
We consider a single-particle tight-binding Hamiltonian for rhombohedral trilayer graphene (RTG) on a Haldane layer (RTG/$\tilde{\rm G}$), as depicted in Fig.~\ref{Fig1}, given by:

\begin{equation}\label{Hamil}
H(\vb{k})=H_{\rm RTG}+H_h+T
\end{equation}

where $H_{\rm RTG}$, $H_h$, and $T$ represent the Hamiltonians of RTG, the Haldane layer, and the interaction between the Haldane layer and the closest layer of RTG, respectively. The RTG Hamiltonian $H_{\rm RTG}$ is presented in Appendix~\ref{app.tri_band}. The Haldane layer Hamiltonian is $H^h(\vb{k})=\vb{d^h(k)\cdot\sigma} + \epsilon^h \cdot \mathbb{I}$, where $\sigma$ are Pauli matrices, $d_x^h=t^{\prime}_0\sum_{i=1}^3 cos(\vb{k\cdot e_i})$, $d_y^h=-t^{\prime}_0\sum_{i=1}^3 sin(\vb{k\cdot e_i})$, $d_z^h=M-2t_2^{\prime} sin(\phi)\sum_{i=1}^3 -(-1)^isin(\vb{k\cdot a_i})$, and $\epsilon_h=2t_2^{\prime} cos(\phi)\sum_{i=1}^3 cos(\vb{k\cdot a_i})$. Here, $M$ is the Semenoff mass (on-site potential), $t_{2}^{\prime}$ is the next-nearest-neighbor (NNN) hopping strength, $\phi$ is the NNN hopping phase, and $t^{\prime}_0$ is the nearest-neighbor (NN) hopping strength. The vectors $\vb{e}_i$ and $\vb{a}_i$ connect nearest-neighbor and next-nearest-neighbor lattice sites, respectively. We consider the Haldane layer to be directly below RTG, interacting only with the lower layer of RTG. Note that while we consider a Haldane model with the same lattice constant as graphene, lattice mismatch in graphene/non-graphene bilayers would lead to a Moiré superlattice in the RTG/$\tilde{\rm G}$ heterostructure~\cite{PhysRevLett.122.016401}. The coupling matrix between the two layers is $T=\begin{bmatrix} t_1^{\prime} & -t_4^{\prime}f(\vb{k}) \\ -t_4^{\prime}f^{\dag}(\vb{k}) & t_1^{\prime} \end{bmatrix}$, where $t_1^{\prime}$ and $t_4^{\prime}$ are the vertical and diagonal hopping strengths between the Haldane layer and the lower layer of RTG, respectively. The in-plane nearest-neighbor structure factor is given by $f(\vb{k})=\sum_{j=1}^3 e^{i\vb{k}\cdot \vb{e_j}}$. 
In the presence of broken time-reversal (TR) and inversion (I) symmetries induced by the Haldane layer, non-vanishing Berry curvatures are generated.
The Berry curvatures within the Brillouin zone (BZ) are calculated using the formula:

\begin{equation}\label{Berry}
\Omega_n^{xy}(\vb{k})=-2\sum_{n'\neq n}Im[\frac{\bra{\psi_n}\frac{\partial H}{\partial k_x} \ket{\psi_{n'}}\bra{\psi_{n'}}\frac{\partial H}{\partial k_y} \ket{\psi_n}}{(E_n-E_{n'})^2}]
\end{equation}

Here, $E_n$ represents the nth band, and the summation is over all other bands ($n^{\prime}\neq n$) at each $\vb{k}$ point. $\ket{\psi_n(\vb{k})}$ denotes the Bloch states. The Chern number~\cite{PhysRevLett.71.3697} of these bands is calculated as $C_n=\frac{1}{2\pi}\int d^2(\Vec{k}) \Omega_n(\Vec{k})$.

\begin{figure}
    \centering    
    \includegraphics[width=\columnwidth]{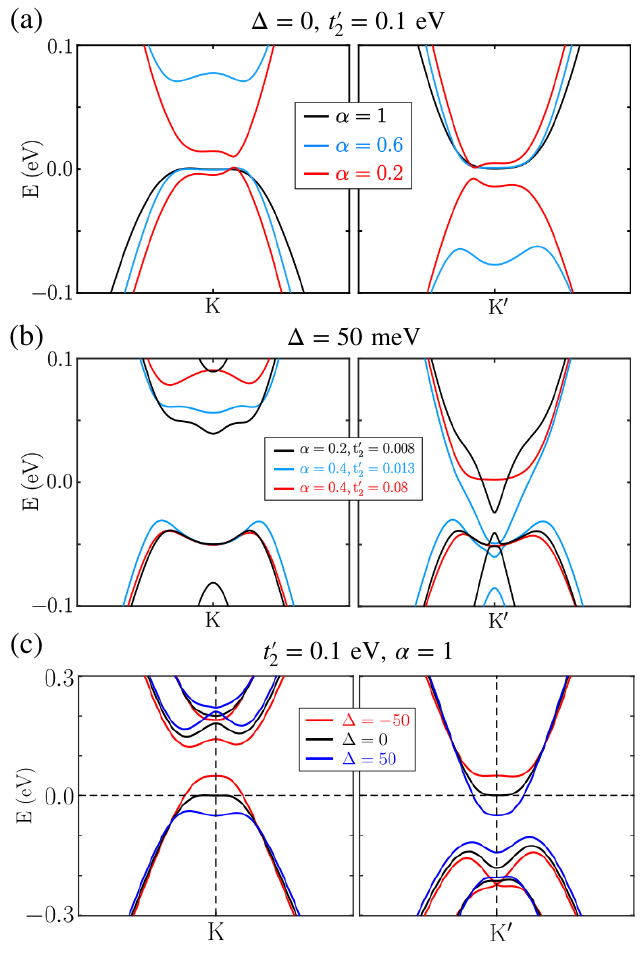}
    \caption{Low energy band structures of RTG/$\tilde{\rm G}$ (a) for different values of $\alpha$ with $\Delta=$ 0 and $t_2^{\prime}=$ 0.1~eV. (b) for different values of $\alpha$ and $t_2^{\prime}$ with $\Delta=$ 50 meV and (c) for different values of $\Delta~(meV)$ with $t_2^{\prime}=$ 0.1~eV and $\alpha=$ 1.}
    \label{Fig2}
\end{figure}
%
\section{\label{sec:3}Results and Discussions}
\subsection{\label{sec:3ab} Proximity Effects and Influence of $\Delta$ and $t_2^{\prime}$}
The low-energy bands of RTG, presented in Appendix~\ref{app.tri_band}, show flat bands around the $\rm K$/$\rm K^{\prime}$ points and a van Hove singularity close to the charge neutrality point. These bands originate from the unperturbed A1 and B3 sublattices \cite{Zhang2010}. To achieve topological proximity, we consider a Haldane model layer ($\tilde{G}$) based on graphene to avoid the lattice mismatch complexity. Under the proximity effects, one of these sublattices (A1 or B3) will get perturbed, and the low energy bands related to those sublattices get pushed to higher energy. Additionally, the NNN phase $\phi$ leads to breaking both the inversion and time reversal symmetry resulting a gap at charge neutrality. 
Eventually, the proximity effects lift the valley degeneracy, i.e., $E(K)\neq E(K^{\prime})$ along with the breaking sublattice symmetry, $E(K)\neq -E(K)$. 
It is further causes to an imbalance carriers at $\rm K$ and $\rm K^{\prime}$ valleys in the BZ. It is clear that, the interlayer coupling strengths between RTG layers ($t_1$) and between RTG and $\tilde{G}$ ($t_1^{\prime}$) are different. We explore the low energy bands with those variations. Fig.~\ref{Fig2}(a) shows the low-energy bands of RTG/$\tilde{G}$ for different values of $\alpha = t_1^{\prime}/t_1$, $t^{\prime}_2 = 0.1$ eV, and $\Delta = 0$. The flat bands near the $\rm K$ arise from the valence band and the flat bands near the $\rm K^{\prime}$ is from conduction band. These flat bands remain intact with varying proximity strength ($\alpha$), while the proximity-perturbed bands move away from the Fermi level as $\alpha$ increases, that is conduction/valanece band at $\rm K$/$\rm K^{\prime}$. In Fig.~\ref{Fig2}(b), we show the effect of $\alpha$ and $t_2^{\prime}$ in low energy bands with $\Delta =$ 50 meV. The variation in the $\alpha$ and $t_2^{\prime}$ has dissimilar effect as discussed above. However, the low energy bands at $\rm K$ valley move away from each other from the charge neutrality. Interestingly, at $\rm K^{\prime}$ valley, the bands gets merged with decreasing value of $\alpha$. The variation of $t_2^{\prime}$ is minimal with the fixed value of $\alpha$. Further, the low-energy bands of $\rm RTG/\tilde{\rm G}$ are strongly affected by the interlayer potential difference ($\Delta$), as shown in Fig.~\ref{Fig2}(c). The applied interlayer potential difference ($\Delta$) causes the low-energy bands to cross the $\rm E_{Fermi} = 0$ at $\rm K$/$\rm K^{\prime}$ with the direction of $-/+~\Delta$. Notably, the sign of $\Delta$ determines whether the valence or conduction band crosses the Fermi level near the $\rm K$ or $\rm K^{\prime}$ points, leads to metallic states and valley imbalanced carriers. 
\begin{figure}
    \centering    
    \includegraphics[width=0.8\columnwidth]{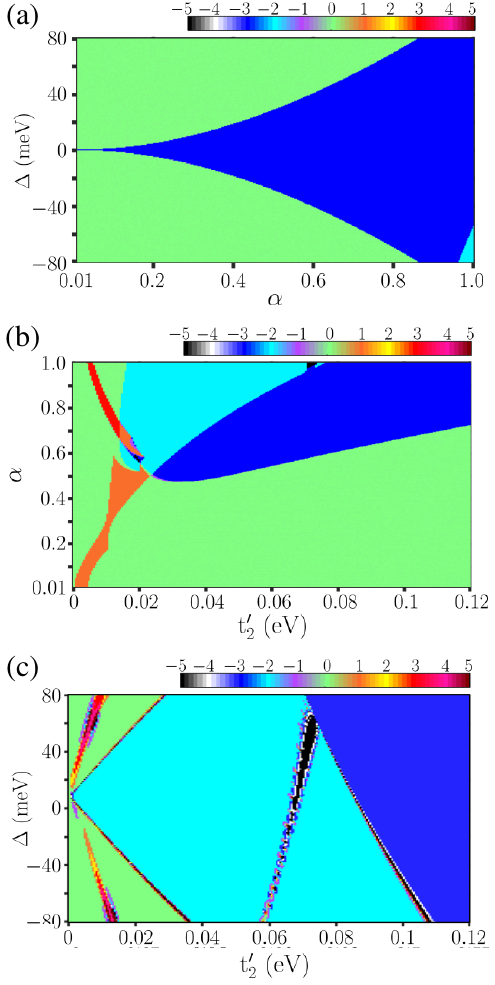}
    \caption{Chern phase diagram of the $\rm{VB1}$ as a function of (a) $\Delta$ and $\alpha$ with $t_2^{\prime}=0.1$ eV, (b) $\alpha$ and $t_2^{\prime}$ with $\Delta =$ 50 meV and (c) $\Delta$ and $t_2^{\prime}$ with $\alpha =$ 1.}
    \label{Fig3}
\end{figure}
\subsection{\label{sec:3ab} Chern Phase diagrams}
The low energy bands are greatly influenced by the choice of $\Delta$, $\alpha$ and $t_2^{\prime}$ as shown in Fig.~\ref{Fig2}. The low-energy flat bands (VB$_1$ and CB$_1$) near the Dirac point are isolated and have non-trivial gaps over the Brillouin zone.  We expect that the topology of these bands can be tuned using the interlayer potential difference ($\Delta$), $\alpha$ and $t_2^{\prime}$. We present the Chern number variation of the valence band (VB$_1$) in Fig.~\ref{Fig3}(a) with $t^{\prime}_2 = 0.1$ eV with a parameter space of proximity strength ($\alpha = 0.01-1.0$) and $\Delta = 0 - \pm 80$ meV. The low-energy bands become trivial for weak $\alpha$ for any value of $\pm \Delta$. It is then noted that, the topological bands with Chern number $C = -3$ is possible with the topological proximity in RTG/$\tilde{G}$ only with the strong $\alpha$ value under the influence of $\Delta$. The Chern number values for the conduction band (CB$_1$) has an opposite sign of the Chern number of VB$_1$.  
Additionally, we show that the topological proximity is influenced by the strength of the NNN hopping ($t^{\prime}_2$). 
\par Fig.~\ref{Fig3}(b) shows the Chern number variation as a function of $\alpha$ and $t^{\prime}_2$ for VB$_1$ with $\Delta = 50$ meV. Topological proximity effectively induces topology in RTG for $\alpha > 0.5$ and $t^{\prime}_2 > 0.02$ eV with Chern number $C = -2$ and shows a topological transition by changing the Chern number to $C = -3$ with the increasing $t^{\prime}_2$. For weaker proximity ($0 < \alpha < 0.5$ and $0 < t^{\prime}_2 < 0.02$), topology emerges with increasing $\alpha$ and $t^{\prime}_2$ having a Chern number $C = 1$. Fig.~\ref{Fig2}(b) shows the low-energy bands for selected values of $\alpha$ and $t^{\prime}_2$ within this weak topology regime. Also, for $0.5 < \alpha < 1.0$ and $0 < t^{\prime}_2 < 0.02$, topology emerges with decreasing $\alpha$ and $t^{\prime}_2$ having a Chern number $C = 3$. Fig.~\ref{Fig7} demonstrates that the low-energy bands remain topological even under zero interlayer potential for $0.01 < \alpha < 1.0$ and $0 < t^{\prime}_2 < 0.12$ eV with Chern number $C = -2$ and shows a topological transition by changing the Chern number to $C = -3$ with the increasing $t^{\prime}_2$. 

\par Fig.~\ref{Fig3}(c) shows the Chern number variation of VB$_1$ as a function of $\Delta$ and $t_2^{\prime}$ with $\alpha=1$. At $t_2^{\prime} = 0$, the low-energy bands are trivial with any interlayer potential difference. 

For $t_2^{\prime} \neq 0$, VB$_1$ becomes topologically non-trivial with Chern numbers for a short range of $\pm \Delta$.
Also, for $0<t_2^{\prime}< 0.02$ eV with increasing $\Delta$, we see unstable topological strips with chern number $C = 2 ~\& ~3$.
For $t_2^{\prime} > 0.04$ eV, VB$_1$ is a topological band with non-zero Chern numbers $C = -2$ for the full range of $\Delta = \pm 80$ meV. 
Interestingly, fragile topological transitions occur for $0.05 < t_2^{\prime} < 0.07$ eV shows a Chern number of $C = -5$, and a stable transition arises with $C = -3$ for $t_2^{\prime} > 0.07$ eV with varying $\Delta$. 
\par From Fig.~\ref{Fig3}, we could see that the low-energy bands Chern number remains intact for $\alpha = 1$ and $t_2^{\prime} > 0.07$ eV. However, the low-energy bands Chern number primarily depend on the relative strengths of $\Delta$ and $\alpha$. 
This indicates that the sufficiently strong proximity strength is necessary for the gate-tunable topological bands in RTG/$\tilde{G}$. So, throughout the manuscript we present results for $\alpha = 1$ and $t^{\prime}_2 = 0.1$ eV unless mentioned. 
\begin{figure}
    \centering    
    \includegraphics[width=0.8\columnwidth]{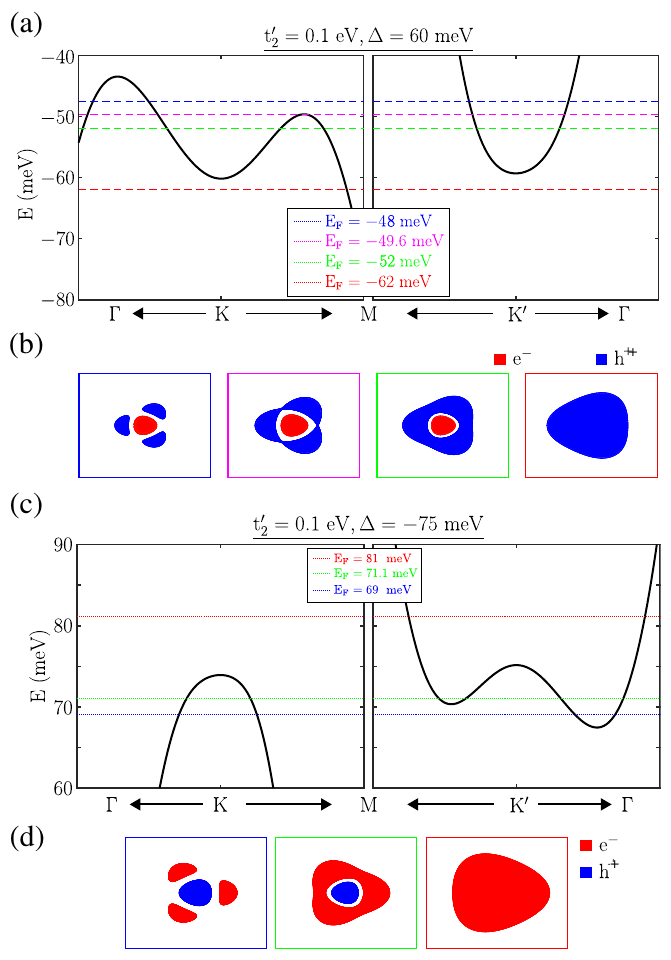}
    \caption{The low energy bands and the Fermi surfaces of RTG/$\tilde{G}$ for applied $\Delta$ and carriers doping. Different possible Fermi surface topology with $e^-$ (red region) and $h^+$ (blue region) carrier fillings for different $E_F$'s along with bands. (a-b) for $\Delta>0$ and (c-d) for $\Delta<0$. }
    \label{FS}
\end{figure}
\subsection{Fermi Surface topology}
Topological proximity in RTG induces unequal carrier populations at the $\rm K$ and $\rm K^{\prime}$ valleys as shown in Fig.~\ref{Fig2}. For applied $\Delta$ and chemical potential $\mu {\rm(meV)}$, the low energy bands cross the Fermi level ($E_F$) leading to various Fermi surface topology. Moreover, the Fermi surface topology further depend on the sign of the $\Delta$. 
In Fig.~\ref{FS}, we show the low energy bands obtained for $t_2^{\prime} = 0.1$ eV with $\Delta>0$ and $\Delta<0$.  
The topological proximity in RTG can induce three types of Fermi surfaces based on $\Delta$ and carrier doping: (i) single, (ii) annular and (iii) pockets shapes. 

In Fig ~\ref{FS} (a), for Fermi level shift $E_F = $ -62 meV, we have Fermi surface of holes ($h^+$) at $K$ valley. Similarly, Fermi surface of electrons ($e^-$) is found for $E_F = $ -81 meV with $\Delta = $ -75 meV locating at $K^{\prime}$ valley. In both the cases, either $e^-$ or $h^+$ dominance lead to have valley imbalance. For the $E_F$ shift close to the charge neutrality, as shown in Fig.~\ref{FS}, both $e^-$ and $h^+$ are having contribution either at $K$/ $K^{\prime}$ leading to form annular Fermi surfaces. For instance, with $\Delta =$ 60 meV for $E_F = $ -49.6 and 52 meV, we have annular Fermi surface with large $h^+$ carriers dominance over the $e^-$ carriers. Also, we have annular Fermi surface with large $e^-$ carriers dominance over the $h^+$ carriers for $\Delta =$ -75 meV. The Fermi surface carriers of type $e^-$ or $h^+$ carriers is denpedent on the direction of the applied inter layer potential difference. In above two cases, we have contribution from  $\rm K$ and $\rm K^{\prime}$.  The pockets shape Fermi surface is found with near equal contribution from $e^-$ and $h^+$ carriers for $E_F = $ -48 and 69 meV with $\Delta =$ 60 and -75 meV, respectively. Interestingly, the Fermi surfaces at $\rm K$ and $\rm K^{\prime}$ with electrons ($e^-$) and holes ($h^+$) carriers are imposed on each other, and found no-cross over between them. 
Recently, the valley imbalance in RTG is explored with spin-orbit proximity ~\cite{Arp2024}. In the case of spin-orbit proximity Fermi surfaces have electrons ($e^-$) and holes ($h^+$) carriers pocket cross over among the valleys lead to inter valley coherence~\cite{Arp2024}. In the topological proximity RTG the annular Fermi surfaces are of tological bands, which may enhance the tendency for unconventional superconductivity through the Kohn-Luttinger mechanism, as discussed in Ref.~\cite{PhysRevB.105.134524}.
\begin{figure}
    \centering
    \includegraphics[width=0.9\columnwidth]{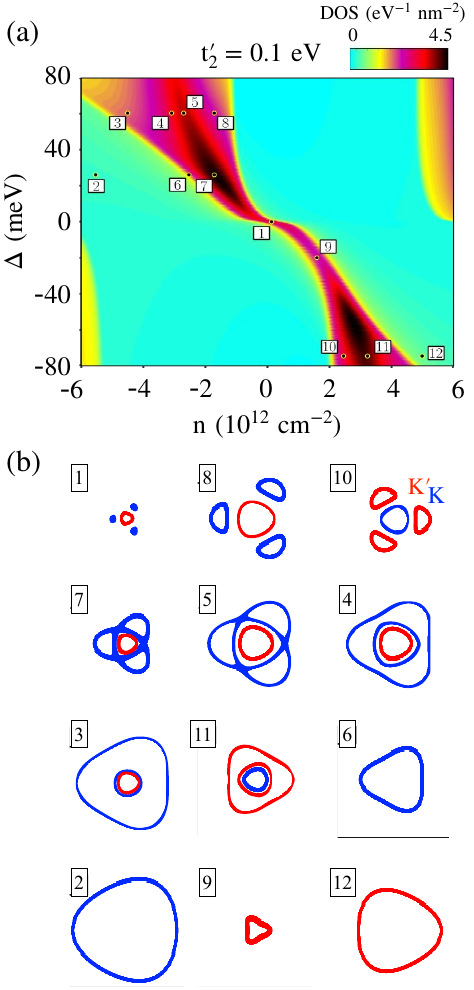}
    \caption{(a) DOS as a function of density ($n$) and interlayer potential difference ($\Delta$) with $t_2^{\prime} = $ 0.1 eV. (b) The Fermi surfaces of the competing states are representing at [$n~(10^{12}~cm^{-2}),\Delta$~(meV)], i.e., (1) [0.16,0], (2) [-5.5,25], (3) [-4.5,60], (4) [-3.1,60], (5) [-2.7,60], (6) [-2.5,25], (7) [-1.6,25], (8) [-1.7,60], (9) [1.65,-20], (10) [2.4,-75], (11) [3.22,-75], (12) [5,-75]  }
    \label{Fig5}
\end{figure}
It is important to note that, the Fermi surface topology of the topological bands of RTG is highly influenced by the $t^{\prime}_2$ and $\Delta$, and obviously on the chemical potential ($\mu~\rm (eV)$). 
\par We further discuss the Fermi surface modification in terms of carriers density ($n$)(electrons and holes) instead $\mu~\rm (eV)$. The  $n>0$ indicates the electrons and $n<0$ indicates the holes as discussed in Appendix.~\ref{app.dos}. We show that the density vs DOS has one-to-one correspondence in terms of chemical potential ($\mu~\rm$) as shown in Fig.~\ref{Fig61}. 
Interestingly, with the applied $\Delta$, these Fermi surfaces demonstrates a Lifshitz transitions~\cite{PhysRevB.107.L121405} and transform to a single Fermi surface at $\rm K$ or $\rm K^{\prime}$~\cite{Arp2024}. As discussed in Ref.~\cite{Arp2024,PhysRevB.105.134524}, similar to the spin-orbit proximity, the Fermi surface topology mainly depends on the carriers doping ($n$) and $\Delta$. 
\par We now discuss the Fermi surface modifications as a function of $n$ and $\Delta$. In Fig.~\ref{Fig5}(a) we show the variation of DOS with simultaneous change in the $n$ and $\Delta$.  At the Charge neutrality we have smaller Fermi surface pockets surrounded at $\rm K$ and a smaller Fermi pocket at the $\rm K^{\prime}$, as shown in Fig.~\ref{Fig5}b (1). For $n =$ -1.6 $\times$ 10$^{12}$ cm$^{-2}$ doping with $\Delta =$ 25 meV, the smaller Fermi surface pockets at charge neutrality enlarged and merges to three-leaf clover-shaped at $\rm K$ and a smaller Fermi pocket at the $\rm K^{\prime}$ (see Fig.~\ref{Fig5}b (7)), which is the demonstration of the Lifshitz transition. A similar shape of Fermi surface, as shown in Fig.\ref{Fig5}b (5), is also observed for the $n =$ -2.7 $\times$ 10$^{12}$ cm$^{-2}$  with $\Delta =$ 60 meV. The Fermi surface for $n =$ 1.65 $\times$ 10$^{12}$ cm$^{-2}$ with $\Delta =$ -20 meV into the conduction band, smaller pocket is located at the $\rm K^{\prime}$. Moreover, in the conduction band side, for the large values of $\Delta = $ -75 meV, the Fermi surface changes drastically with the increasing value of electron doping, as highlighted in Fig.\ref{Fig5}b (10), (11) and (12). For $n =$ 2.4 $\times$ 10$^{12}$ cm$^{-2}$, the Fermi surface is similar like at Charge neutrality. The holes carries, $n =$ -1.7 $\times$ 10$^{12}$ cm$^{-2}$, close to the charge neutrality with large $\Delta =$ 60 meV has similar Fermi surface as at Charge neutrality.   For $n =$ 3.22 $\times$ 10$^{12}$ cm$^{-2}$ the Fermi surface is annular at $\rm K^{\prime}$ and a smaller pocket at $\rm K$. In the valence band side, there are annular Fermi surfaces are possible for the $n =$ -4.5 $\times$ 10$^{12}$ cm$^{-2}$ and $n =$ -3.1 $\times$ 10$^{12}$ cm$^{-2}$ for the $\Delta = $ 60 meV. However, the annular Fermi surface is now at $\rm K^{\prime}$ in electron doping. Interestingly, for $n =$ 5 $\times$ 10$^{12}$ cm$^{-2}$ with $\Delta =$ -75 meV, it is a single Fermi surface at $\rm K^{\prime}$ for the conduction band side. A similar single Fermi surfaces are arise for  $n =$ -5.5 $\times$ 10$^{12}$ cm$^{-2}$ and $n =$ -2.5 $\times$ 10$^{12}$ cm$^{-2}$ with $\Delta =$ 25 meV, as shown in Fig.\ref{Fig5}b (2), (6). From the above discussion, in topological proximity RTG the carrier doping drives the formation of valley-specific Fermi surfaces, and the polarity of the electric fields ($\Delta$) determines the dominance of specific carriers.

\begin{figure}
    \centering    
    \includegraphics[width=0.8\columnwidth]{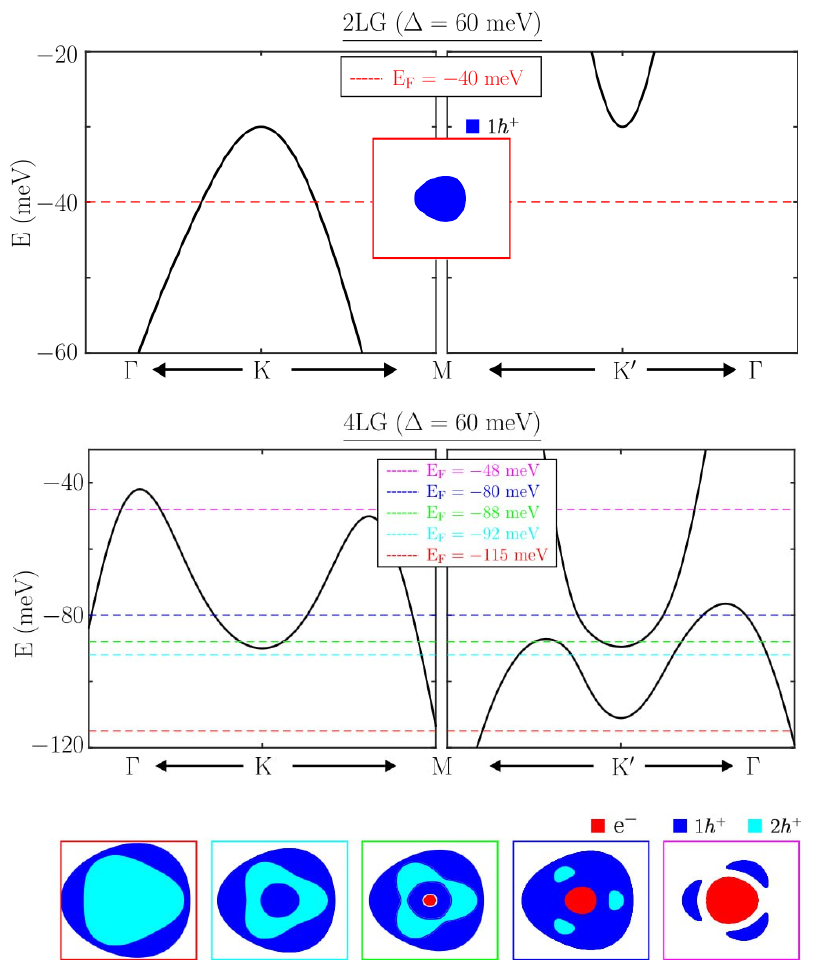}\\
    \caption{The low energy bands and the Fermi surfaces of bilayer RG/$\tilde{G}$ and tetralayer RG/$\tilde{G}$ for applied $\Delta =$60 meV and carriers doping. Different possible Fermi surface topology with $e^-$ (red region) and $h^+$ (blue region) carrier fillings for different $E_F$'s along with bands.}
    \label{FS2and4}
\end{figure}

\section{Multilayer RG/$\tilde {G}$}

The shape and carriers of Fermi surface have a direct influence in the carrier transport properties. In this section, we show a similar Fermi surface formation in case of multilayer RG/$\tilde {G}$. In the Fig.~\ref{FS2and4}, we discuss the cases of 2LG/$\tilde {G}$ and 4LG/$\tilde {G}$. In 2LG/$\tilde {G}$, for the $\Delta =$ 60 meV with $E_F$ = -40 meV, we have a single Fermi surface of holes ($h^+$) locates at $K$. Also, from the low energy bands (see Fig.~\ref{FS2and4}), by tuning the carriers concentration one can achieve a Fermi surface either of holes ($h^+$) or electrons ($e^-$). In case of 3LG/$\tilde {G}$, the Fermi surfaces have holes ($h^+$) and electrons ($e^-$) together with an imbalance concentration of individual. 
\par Interestingly, in case of 4LG/$\tilde {G}$, the Fermi surface carriers of holes ($h^+$) and electrons ($e^-$) arise at $K^{\prime}$ and holes ($h^+$) alone at $K$. The holes ($h^+$) alone contribute from at both the valleys are of different Fermi velocities for the $\Delta =$ 60 meV with $E_F$ = -115 and -92 meV. Here, we have concentric Fermi surfaces of holes ($h^+$) carriers. These concentric Fermi surfaces usually enhances the electronic transport properties due to increased  mean free path and efficient electron flow by minimizing scattering events, which leads to improved conductivity. At $E_F$ = -88 meV, we see three concentric Fermi surfaces with two of them with carriers of holes ($h^+$) and a small portion with carriers of electrons ($e^-$). The Fermi surfaces further evolve by the Lifshitz transformations for $E_F$ = -80 and -48 meV. At $E_F$ = -48 meV, the Fermi surface has dominance of electrons ($e^-$). In all the cases the opposite carriers are not equal. Thus, we conclude that, the topological proximity in multilayer rhombohedral graphene emerge the valley imbalance.      

\section{Orbital magnetization}
\begin{figure}
    \centering    
    \includegraphics[width=\columnwidth]{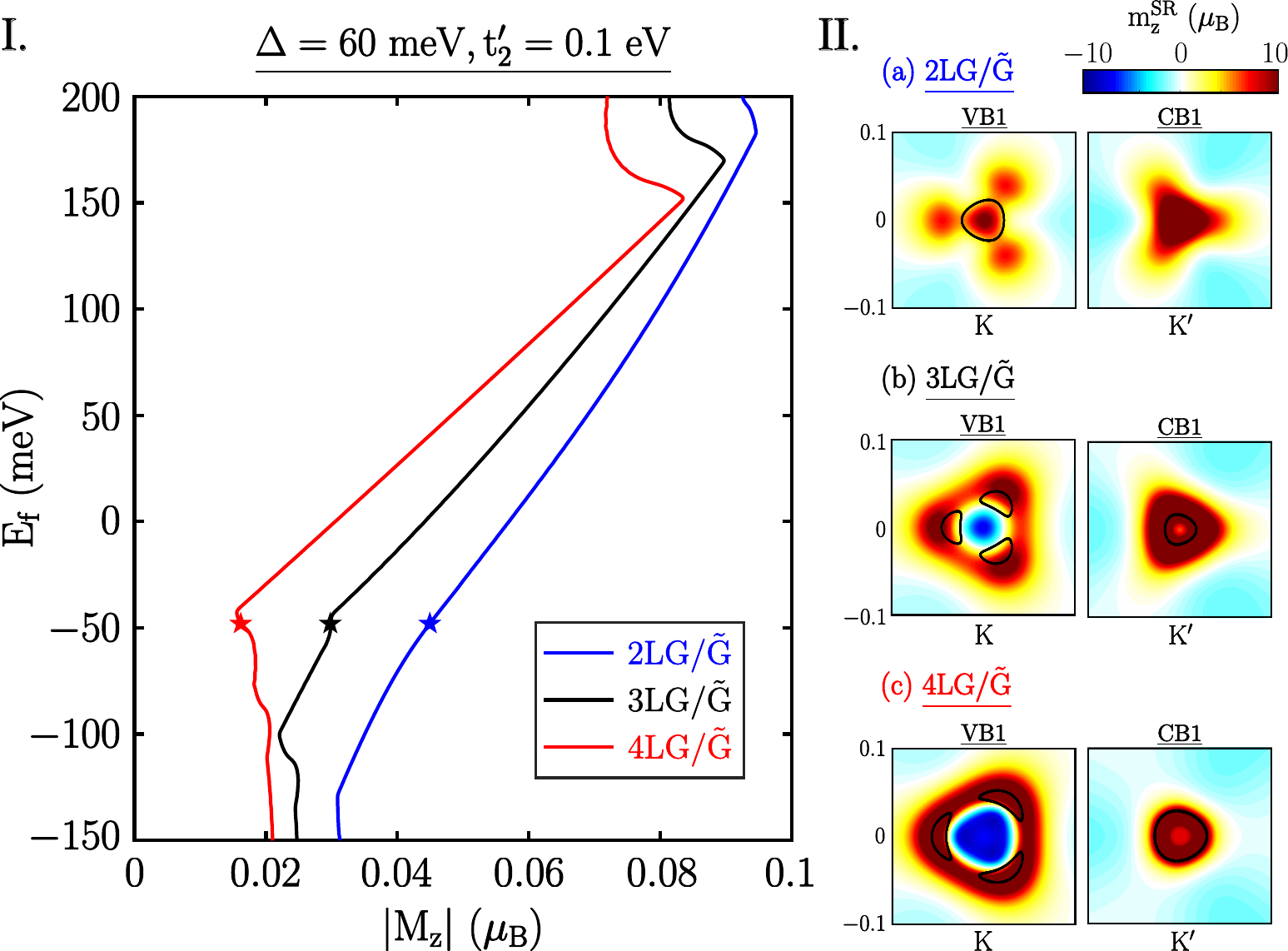}
    \caption{(I). Orbital magnetization of multilayer RG/$\tilde {G}$ as a function of Fermi energy. (II) The corresponding magnetic moment distribution in BZ corners ($\rm K$ and $\rm K^{\prime}$) of $\rm VB1$ and $\rm CB1$. The Fermi surfaces are shown as black surface plots within the magnetic moment distribution for a Fermi level of -48 meV, also labeled as a star in (I).}
    \label{OrbMag}
\end{figure}
The orbital magnetization of these isolated bands arise from the Berry curvature, $\Omega_n^{xy}$, and the topological nature of the bands,
is estimated from the equation  
$M_z^{orb}=-\frac{e}{\hbar} \int \frac{d^2k}{(2\pi)^2}\sum_n f_n(k) \sum_{n^{\prime}\neq n}[(E_n-E_{n^{\prime}})+2(\mu-E_n)]\times Im[\frac{\bra{\psi_n}\frac{\partial H}{\partial k_x}\ket{\psi_{n^{\prime}}}\bra{\psi_{n^{\prime}}}\frac{\partial H}{\partial k_y}\ket{\psi_n}}{(E_n-E_{n^{\prime}})^2}] $.
The calculated $M_z^{orb}$ for 2LG, 3LG and 4LG is presented in Fig.~\ref{OrbMag}(I) for $\Delta =$ 60 meV and $t_2^{\prime} =$ 0.1 eV. We limit the orbital magnetization calculation to the low energy range such that only includes the valence and conduction bands close to the charge neutrality point. Along with the $M_z^{orb}$, we also show the orbital magnetic moment ($m_z^{SR}$) due to the self-rotation of electrons in the orbital. Interestingly, the carriers at $E_f = -$48 meV show a large variation of orbital magnetic moment between $\pm 10~\mu_B$ for all 2L, 3L and 4L RG/$\tilde{G}$. The Fermi surface at the $E_f =-$48 meV (see Fig.~\ref{OrbMag}(II)) indicates the filled carriers included within the Fermi surface in case of conduction band (CB) and excluded from the Fermi surface in case of valence band (VB). Notably, the $M_z^{orb}$ decreases with an increasing number of layers, due to the enhanced mixing of magnetic moment polarity in the valley K, within the low energy range, as shown in Fig.\ref{OrbMag}(II). The magnitude of the orbital magnetization of multilayer RG/$\tilde {G}$ is showing a increased behavior from $|M_z^{orb}|=$ 0.02$\mu_B$ to 0.1$\mu_B$ as function of $E_f$, indicates the accumulation of similar polarity orbital magnetic moment. 
As there is no global band gap in multilayer RG/$\tilde{G}$ (see Fig.\ref{FS}(a) and Fig.\ref{FS2and4}), the slope of the orbital magnetization is not directly proportional to the Chern number ($\frac{dM_z}{dE_f}\neq Ce/h$)~\cite{PhysRevB.107.L121405}. Due to the dominance of the valley specific band gap, orbital magnetization increases linearly beyond a certain $E_f$, as shown in Fig.\ref{OrbMag}(I) Additionally, the slope of $M_z^{orb}$ increases with the number of layers, corresponding to the increasing Chern number.

\section{Conclusion}
We investigated the impact of topological proximity on the electronic structure, Chern numbers and Fermi surfaces of rhombohedral trilayer graphene (RTG). We show that the next-nearest-neighbor hopping ($t^{\prime}_2$) in the Haldane layer significantly alters the low-energy bands of RTG aligned on a Haldane layer ($\tilde{\rm G}$). These proximity effects lift the valley degeneracy and, along with broken sublattice symmetry, lead to an imbalance carriers at $\rm K$ and $\rm K^{\prime}$ valleys in the BZ. We demonstrated that the low energy flat band near the $\rm K$ arise from the valence band and the flat band near the $\rm K^{\prime}$ is from conduction band. However, under the influence of the vertical electric field, these bands evolve into valley-resolved metallic states. Importantly, these metallic states in RTG/$\tilde{\rm G}$ exhibit a non-trivial gap at charge neutrality near the $\rm K$/$\rm K^{\prime}$ points, and the low energy bands become non trivial with Chern numbers of $\pm 3$ for $t^{\prime}_2 = 0.1$ eV. An interlayer potential difference ($\Delta$) induced by a vertical electric field triggers topological transitions among the low-energy non-trivial bands for $t^{\prime}_2 = 0.1$ eV. Furthermore, we explore Fermi surface reconstruction as a function of interlayer potential difference and carrier density. The Fermi surface topology in RTG/$\tilde{\rm G}$ is significantly influenced by the next-nearest-neighbor hopping strength ($t^{\prime}_2$), interlayer potential difference, and carrier doping. Carrier doping drives the formation of valley-specific Fermi surfaces, while the polarity of the electric fields determines the dominance of specific carriers, and lead to non-zero orbital magnetization. We conclude that topological proximity in multilayer rhombohedral graphene leads to valley imbalance and diverse Fermi surface topologies.
\section*{Acknowledgments}
We acknowledge the support provided by the Kepler Computing Facility, maintained by the Department of Physical Sciences, IISER Kolkata, for various computational requirements. S.G. acknowledges the support from the Council of Scientific and Industrial Research (CSIR), India, for the doctoral fellowship. B. L. C acknowledges the SERB with Grant No. SRG/2022/001102 and ``IISER Kolkata Start-up-Grant" No. IISER-K/DoRD/SUG/BC/2021-22/376.

\appendix

\section{Low-energy Bands in $\rm RTG$ }\label{app.tri_band}
\setcounter{figure}{0}
\renewcommand{\thefigure}{A\arabic{figure}}
\begin{figure}
    \centering
    \includegraphics[width=0.5\textwidth]{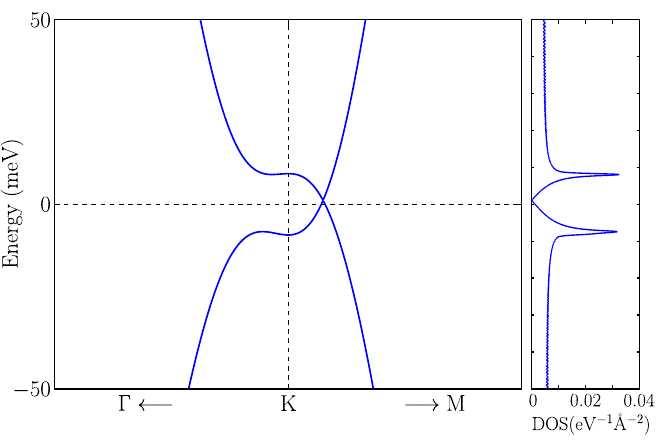}
    \caption{Low energy band dispersion of the rhombohedral stacked trilayer Graphene (RTG) along symmetry path $\Gamma$ - K - M point (left panel) and the corresponding DOS in the right panel.}
    \label{Fig6}
\end{figure}
The Hamiltonian of the rhombohedral tri-layer graphene ($\rm RTG$) in momentum space can be demonstrated in sublattice basis ($A_1,B_1,A_2,B_2,A_3,B_3$) as
\begin{equation}
H_{RTG}(\vb{k})=
\begin{pmatrix}
    H_{11} & H_{12} & H_{13}  \\
    H_{12}^{\dag} & H_{22} & H_{12} \\
    H_{13}^{\dag} & H_{12}^{\dag} & H_{33} 
\end{pmatrix}
\end{equation}
where the intralayer $(H_{ll})_{2\times2}(l=1,2,3)$ and interlayer $(H_{ij})_{2\times2}(i\neq j=1,2,3)$ Hamiltonian terms are given by
\begin{equation}\nonumber
    H_{ll}(\vb{k})=
    \begin{pmatrix}
        u_{A_l} & t_0 f(\vb{k}) \\
        t_0 f^{\dag}(\vb{k}) & u_{B_l}
    \end{pmatrix}+ V_{ll} \mathbb{I}
\end{equation}
\begin{equation}
    H_{12}=\begin{pmatrix}
        t_4 f(\vb{k}) & t_3 f^{\dag}(\vb{k}) \\ t_1 & t_4 f(\vb{k})
    \end{pmatrix}
\end{equation}
\begin{equation}\nonumber
    H_{13}=\begin{pmatrix}
        0 & t_2 \\ 0 & 0
    \end{pmatrix}
\end{equation}
 The $V_{ll}\mathbb{I}$ term includes the inter-layer potential difference due to a applied vertical electric field through $V=\Delta (1,0,-1)$, where we consider equal-magnitude potential drops ($\Delta$) across the consecutive RTG layers. The hopping parameters, namely, $(t_0,t_1,t_2,t_3,t_4)$ are used as (-3.1,0.3561,-0.0083,0.293,0.144) eV, similar to the literature's\cite{PhysRevB.108.155406,Liu2023} that are corresponding to the nearest neighbour intra-layer hopping terms between $A_l$ and $B_l$ and inter-layer $B_l$ and $A_{l+1}$, $A_l$ and $B_{l+2}$, $A_l$ and $B_{l+1}$ and $A_l(B_l)$ and $A_{l+1}(B_{l+1})$ sites respectively. The diagonal-site potentials $u_{A_l}(u_{B_l})$ at each sublattice are $u_{A_1}=u_{B_3}=0$, $u_{B_1}=u_{A_3}=0.0122$ eV and $u_{A_2/B_2}=-0.0164$ eV. The low energy band structure and the corresponding density of state (DOS) of the RTG are shown in Fig.\ref{Fig6}.
 
\section{Chern Phase diagram with $\alpha$ and $t_2^{\prime}$ for $\Delta =$ 0 }\label{app.alpha}
\setcounter{figure}{0}
\renewcommand{\thefigure}{B\arabic{figure}}
\begin{figure}
    \centering    
    \includegraphics[width=0.5\textwidth]{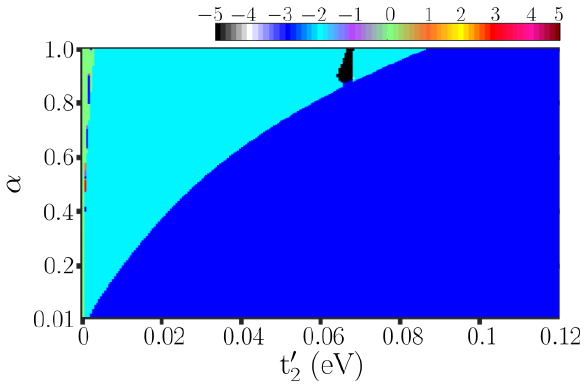}
    \caption{Chern numbers of the $VB_1$ are varies as a function of proximity strength ($\alpha$) and Haldane strength ($t_2^{\prime}$), where interlayer potential difference ($\Delta=0$).}
    \label{Fig7}
\end{figure}
The low-energy Chern bands also show the effect of the proximity strength ($\alpha$) along with the NNN strength ($t_2^{\prime}$) in the absence of the interlayer potential difference $\Delta$. In Fig.~\ref{Fig7}, we presented the VB$_1$ Chern number variation as function of $\alpha$ and $t_2^{\prime}$ with $\Delta =$ 0. The VB$_1$ is trivial with zero Chern number for $t_2^{\prime}=$ 0 eV, and for non-zero values of $t_2^{\prime}$, VB$_1$ shows a Chern number of -2. Further, the VB$_1$ acquire a Chern number of -3 for a value of $\alpha = 0.01$, and remains constant with the increasing value of $t_2^{\prime}$. Moreover, the Chern number changed to -2 with the increasing value of $\alpha$ for $0<t_2^{\prime}<0.08$~eV. It indicates that, the topological phase transition is happened with increasing value of $t_2^{\prime}$ for higher values of $\alpha$.
\section{DOS vs Density}\label{app.dos}
\setcounter{figure}{0}
\renewcommand{\thefigure}{C\arabic{figure}}
\begin{figure}[h!]
    \centering
    \includegraphics[width=0.5\textwidth]{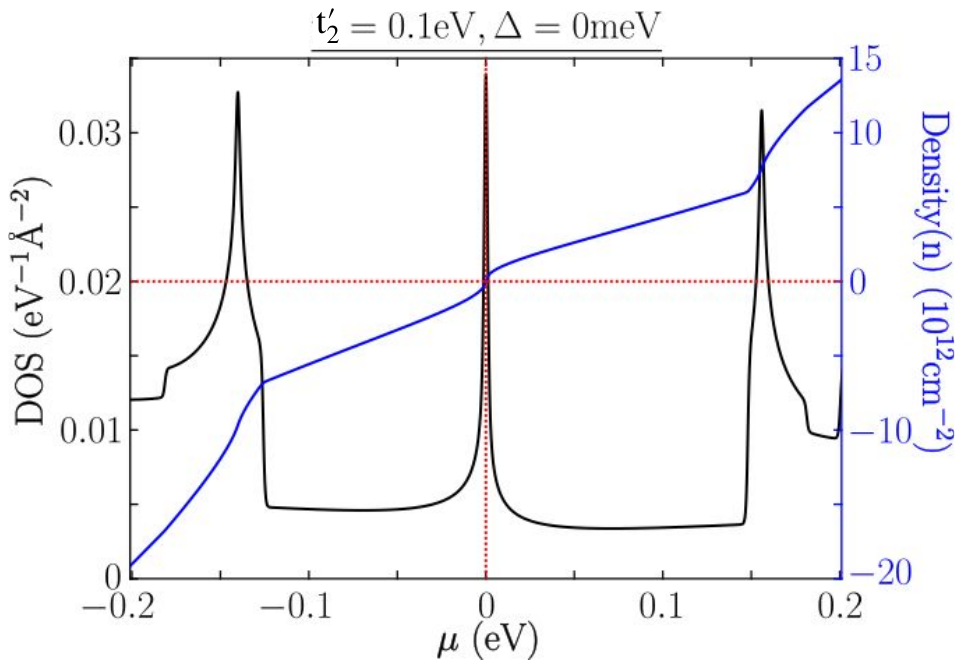}
    \caption{The DOS and Density ($n$) as a function of chemical potential ($\mu$ eV) for $t_2^{\prime} = 0.1$ eV and $\Delta = 0$.}
    \label{Fig61}
\end{figure}
We investigated the dependence of the low-energy bands on the interlayer potential difference ($\Delta$), we calculated the density of states (DOS) $\nu(\mu)$ and charge density $n(\mu)$ as functions of $\Delta$ and chemical potential $\mu$. For fixed values of $\Delta$, we numerically diagonalized the Hamiltonian (Eq.~\ref{Hamil}) to obtain the single-particle energies at each point of the momentum grid. The functions $n(\mu)$ and $\nu(\mu)$ were then calculated by summing the Fermi-Dirac distribution $n_F(\epsilon_k-\mu)$ and its derivative $n_F^{\prime}(\epsilon_k-\mu)$  over all points in the grid, as described in Ref.~\cite{PhysRevLett.121.167601}. The relevant equations are:
\begin{equation}{\label{density_eq}}
    n(\mu) = 2g_{sym}\frac{S_k}{(2\pi)^2}\frac{1}{N}\sum_k n_F(\epsilon_k-\mu)
\end{equation}
\begin{equation}{\label{dos_eq}}
    \nu(\mu) = 2g_{sym}\frac{S_k}{(2\pi)^2}\frac{1}{N}\sum_k n_F^{\prime}(\epsilon_k-\mu)
\end{equation}
Here, $N=\sum_k ~1$ represents the total number of momentum points within the Brillouin zone, with an area of $S_k$. The factor of 2 accounts for the spin degeneracy. The symmetry factor $g_{sym} = 1$ reflects the whole Brillouin zone. The normalization constant in Equation~\ref{density_eq} and \ref{dos_eq} is chosen to ensure that $n(\mu)$ and $\nu(\mu)$ have the correct physical units of $\rm eV^{-1}\AA^{-2}$ and $\rm 10^{12}cm^{-2}$, respectively. We show the one-to-one mapping of DOS and density in \ref{Fig61}.
\bibliography{sample}
\end{document}